\documentclass[a4paper,14pt]{extarticle}
\usepackage{cmap} 
\usepackage[cp1251]{inputenc}
\usepackage[english,russian]{babel}
\usepackage{braket,amssymb,amsmath,array,commath,mathtext,wrapfig,tikz}
\usepackage[arrows=pgf-filled,version=3]{mhchem}
\usepackage{bm}
\usepackage{amsthm,amsfonts,amscd}
\usepackage{graphicx}
\usepackage{cite}
\usepackage{xcolor}
\usepackage{lscape}
\usepackage{geometry}
\usepackage{indentfirst}
\usepackage{afterpage,soul,ulem,dcolumn,changebar,longtable,hhline,multirow,makecell}
\usepackage[small,bf]{caption}
\makeatletter
\renewcommand{\@biblabel}[1]{#1.\hfil} 
\makeatother

\begin{document}
\renewcommand{\refname}{References}
\begin{center}
\textit{On the particle collisions during gravitational collapse of Vaidya spacetimes}
\end{center}
	\begin{center}
\textbf{Vitalii Vertogradov}\\
	\end{center}

	\begin{center}
		Physics department, Herzen state Pedagogical University of Russia,
		
		48 Moika Emb., Saint Petersburg 191186, Russia
		
		SPb branch of SAO RAS, 65 Pulkovskoe Rd, Saint Petersburg 196140, Russia
		
		vdvertogradov@gmail.com
	\end{center}
	
\textbf{Summary:} The center-of-mass energy can be arbitrarily high in Schwarzschild spacetime if one considers the front collision of two particles, one of which moves along so-called white hole geodesics and another one along a black hole geodesic. This process can take place if one considers the gravitational collapse model. In this paper, we consider the well-known naked singularity formation in Vaidya spacetime and investigate the question about two particle collision near the boundary of the collapsing cloud. The center-of-mass energy of the front collision is considered. One particle moves away from the naked singularity and another one falls onto a collapsing cloud. We show that the center-of-mass energy grows unboundly if the collision takes place in the vicinity of the conformal Killing horizon. 

\textbf{Key words}: Gravitational collapse, Particles collision, Vaidya spacetime, Naked singularity, Conformal symmetry

\section{introduction}

The center-of-mass energy of two particle collision can grow unboundly in the Kerr spacetime~\cite{bib:bsw} if one of the particles is fine-tunned (so-called critical particle~\cite{bib:zaslavtan}). This effect was firstly proposed by Ba~ {n} ados, Silk and West and is called BSW effect. Original version of this effect declares absence of the unbound energies in Schwarzschild and Reissner-Nordstrom spacetimes. However, it was shown that this effect is possible in Reissner-Nordstrom-anti- de Sitter spacetime~\cite{bib:zaslavanti}. Despite the unbound the center-of-mass energy, a distant observer will measure small amount of the energy due to this process in the Kerr spacetime~\cite{bib:japan} and an escaping particle will be able to carry away arbitrarily large amount of energy in. Reissner-Nordstrom case~\cite{bib:zaslavextract}.

In spite the fact that BSW effect is absent in Schwarzschild spacetime, one  can still obtain the unbound center-of-mass energy of two colliding particles~\cite{bib:pavlov}. Due to geodesic completeness, there must be geodesics which appear in our Universe from the region inside the gravitational radius i.e. so-called white hole geodesics. For example, geodesics for particles with negative energy in the Kerr metric are such geodesics~\cite{bib:grib, bib:vertog}. One can imagine the following situation: the first particle moves along the white hole geodesic away from the gravitational radius and the second particle moves along black hole geodesics falling onto a black hole. As the result, one can observe the front collision in the vicinity of the event horizon and due to this process the center-of-mass energy can grow unboundly. The problem is that the Schwarzschild black hole is an eternal one and if one follows the geodesic back then it must appear from the collapsing cloud. So to understand the front collision in the vicinity of the event horizon in Schwarzschild spacetime, one must consider the gravitational collapse problem. The nature of white hole geodesics can be explained by the naked singularity formation due to gravitational collapse problem~\cite{bib:vernature}. The outcome of the gravitational collapse might be not only a black hole but also a naked singularity~\cite{bib:joshi1, bib:joshi2, bib:dey}. The naked singularity formation in Vaidya spacetime has been considered in~\cite{bib:dvedy}. The gravitational collapse of the generalized Vaidya spacetime and the naked singularity formation has been investigated in~\cite{bib:mah, bib:ver1, bib:ver2, bib:ver3}. In the case of the eternal naked singularity formation in Vaidya spacetime~\cite{bib:ver1, bib:ver4} the unbound center-of-mass energy is possible only in the vicinity of the singularity. 

In this paper, we consider the following model: a particle moves along the non-spacelike, future-directed  geodesic which terminates at the naked singularity in the past. When the apparent horizon forms, the particle is in the vicinity of the apparent horizon and outside it. At this time, the second particle, moving along a black hole geodesic, falls onto a black hole. As the result we have the front collision of two particles. We estimate the center-of-mass energy of this process and find out where this process should take place to have an unbound energy collision.

This paper is organized as follows: in sec. II we consider a well-known gravitational collapse model of Vaidya spacetime and show the naked singularity formation. We also show that the geodesics can originate at this singularity. In sec. III we introduce the coordinate transformation and consider Vaidya spacetime in conformally static coordinates. In this case we investigate the center-of-mass energy of the two particle front collision.  Sec. IV is the conclusion.

The system of units $G=c=1$ will be used throughout the paper. We use the signature $-\,, +\,, +\,, +$.

\section{The naked singularity formation in Vaidya spacetime}

The Vaidya metric~\cite{bib:vay} describes a dynamical spacetime instead of a static spacetime as the Schwarzschild or Reissner-Nordstrom  metrics do. In the real world, astronomical bodies gain mass when they absorb radiation and they lose mass when they emit radiation, which means that the space-time around them is time-dependent. Papapetroo~\cite{bib:pap} showed that Vaidya spacetime violates the cosmic censorship conjecture and contains the naked singularity. The line element in Eddington-Finkelstein coordinates has the following form:
\begin{equation} \label{eq:metric}
ds^2=-\left(1-\frac{2M(v)}{r}\right)dv^2+2dvdr+r^2d\omega^2 \,.
\end{equation}

Here $M(v)$ the time-depended mass of a black hole, $d\Omega^2=d \theta^2+\sin^2\theta d\varphi^2$ is the metric on unit sphere.

The apparent horizon equation is given by~\cite{bib:pois}:
\begin{equation}
r_{ah}=2M(v) \,.
\end{equation}

The first shell collapses at $r=0$ at the time $v=0$ and the singularity forms at this time. The singularity is naked if at the time of the singularity formation $v=0$ the apparent horizon doesn't form and there is a family of non-spacelike, future-directed geodesics which terminate at the central singularity in the past. Let's prove the last statement. For this purpose, we define the mass function as:
\begin{equation} \label{eq:lambda}
M(v)=\mu v \,, \mu>0 \,.
\end{equation}
Here $\mu$ is a positive constant. Here, we just show that naked singularity is possible. The thorough investigation of this model one can find at~\cite{bib:dvedy}.

To prove the existence of a family of non-spacelike, future-directed geodesics which terminate at the central singularity in the past, one should consider the null radial geodesic, which, for the metric \eqref{eq:metric} with the mass condition \eqref{eq:lambda} has the following form:
\begin{equation} \label{eq:geod}
\frac{dv}{dr}=\frac{2r}{r-2\mu v} \,.
\end{equation}

The solution $v=const.$ doesn't suit us because of infalling matter $v=const.$ corresponds to ingoing geodesics and we are interested in outgoing ones. So, \eqref{eq:geod} corresponds to outgoing geodesic if the following condition is held:
\begin{equation}
\lim\limits_{v\rightarrow 0\,, r\rightarrow 0} \frac{dv}{dr}=X_0>0 \,.
\end{equation}

If the value $X_0$ is positive and finite than the geodesic \eqref{eq:geod} is outgoing one. Let's consider the limit in \eqref{eq:geod}:
\begin{equation}
X_0=\frac{2}{1-2\mu X_0} \,.
\end{equation}

From this equation we obtain:
\begin{equation} \label{eq:root}
X^{\pm}_0=\frac{1\pm\sqrt{1-16\mu}}{4\mu} \,.
\end{equation}
From this equation, one can see that in the case of the linear mass function and  if $\mu<\frac{1}{16}$ then the outcome of the gravitational collapse might be  the naked singularity formation. Hence, there might be particles which move away the singularity and now we are ready to consider the front collision of two particles.

\section{The front collision Effect in Vaidya spacetime}

The Vaidya spacetime \eqref{eq:metric} is time-depended and because of it one has only one conserved quantity - the angular momentum $L$. In the general case, the Vaidya spacetime doesn't possess any additional symmetry. However, for the particular choice of the mass function, the metric \eqref{eq:metric} admits the conformal Killing vector~\cite{bib:mahconformal}. In this case $M$ must  have the following form~\cite{bib:nelson}:
\begin{equation} \label{eq:mass}
M(v)=\mu v \,, \mu >0 \,.
\end{equation}
Where $\mu$ is the positive constant. As we found out in the previous section if $\mu<\frac{1}{16}$, the gravitational collapse might end with the naked singularity formation. Further, in the paper, we impose the condition $\mu <\frac{1}{16}$ because we are interested in temporal naked singularity formation. If we take into account this condition and \eqref{eq:mass} then by coordinate transformation~\cite{bib:german}:
\begin{equation}
\begin{split}
v=r_{0}e^{\frac{t}{r_0}} \,, \\
r=Re^{\frac{t}{r_0}} \,.
\end{split}
\end{equation}
one obtains Vaidya metric in conformally static coordinates:

\begin{equation} 
\label{eq:metric2}
ds^2=e^{\frac{2t}{r_0}}\left[ -\left(1-\frac{2\mu r_{0}}{R}-\frac{2R}{r_0}\right)dt^2+2dtdR+R^2 d\Omega^2 \right]\,.
\end{equation}

We will consider the movement in the equatorial plane $\theta=\frac{\pi}{2}$. The metric \eqref{eq:metric2} admits the conformal Killing vector $\frac{d}{dt}$, which is timelike in the region:
\begin{equation} \label{eq:region}
1-\frac{2\mu r_0}{R}-\frac{2R}{r_0}>0 \,.
\end{equation}
It means that one has the conserved energy along null geodesics in the region \eqref{eq:region} i.e.:
\begin{equation} \label{eq:energy}
E=e^{\frac{2t}{r_0}}\left(1-\frac{2\mu r_0}{R}-\frac{2R}{r_0}\right)\frac{dt}{d\lambda}-e^{\frac{2t}{r_0}}\frac{dR}{d\lambda}  \,.
\end{equation} 
The angular momentum $L$ has the following form:
\begin{equation} \label{eq:momentum}
L=e^{\frac{2t}{r_0}}R^2\frac{d\varphi}{d\lambda} \,.
\end{equation}
The problem is that the energy \eqref{eq:energy} is not a constant of motion along timelike geodesics. Fortunately, the conformal Killing vector $\frac{d}{dt}$ is the homothetic Killing vector~\cite{bib:newbook} and this fact allows us to find another constant of motion along timelike geodesic:
\begin{equation} \label{constantofmotion}
\varepsilon =E-\lambda \,.
\end{equation}
Despite this quantity \eqref{eq:constantofmotion} depends on the affine parameter $\lambda$, it is conserved charge along any geodesics.

To find the $\frac{dR}{d\lambda}$ component of the four velocity $u^i$, one should substitute \eqref{eq:constantofmotion} and \eqref{eq:momentum} into the timelike condition $g_{ik}u^iu^k=-1$. One obtains:
\begin{equation} \label{eq:radial}
e^{\frac{4t}{r_0}}\left(\frac{dR}{d\lambda}\right )^2=E^2-e^{\frac{2t}{r_0}}\left(1-\frac{2\mu r_0}{R}-\frac{2R}{r_0}\right)\left(\frac{L^2}{r^2e^{\frac{2t}{r_0}}}+1\right)=e^{\frac{4t}{r_0}}P^2_R \,.
\end{equation}

Where $P_R=P_R(R, t)$ is some positive function. According to BSW effect~\cite{bib:bsw}, the energy of the centre of mass $E_{c.m.}$ of two colliding particles for the extremal Kerr black hole can grow unboundly. For this purpose, one of the particles must be critical one~\cite{bib:zaslavtan}. In Schwarzschild spacetime, the energy $E_{c.m.}$ is finite according to the original proposal. However, if we consider, in Schwarzschild spacetime, the collision of two particles one of which moves along white hole geodesic from the gravitational radius and another one moves along a black hole geodesic and fall onto a black hole~\cite{bib:pavlov} then the energy $E_{c.m.}$ of the collision can be unbound. The problem is that the Schwarzschild metric describes the eternal black hole and the white hole geodesic appears from the region outside the white hole gravitational radius in past infinity. However, if one considers the physically relevant model, then prolonging the white hole geodesic into the past, one can see that it appears from the collapsing cloud of the matter. So to understand this analogy of  the BSW effect one, first of all, should consider the gravitational collapse model which, in the case of Vaidya spacetime, has been done in the previous section. We have proven that if $\mu<\frac{1}{16}$ the gravitational collapse might end with the naked singularity formation. For our model it means that there is a family of non-spacelike future-directed geodesics which terminate at the central singularity in the past. Let's consider the following situation: the particle $1$ moves along such geodesic and when the apparent horizon forms, this particle is in the vicinity of this horizon  and outside it. At this time the particle $2$, which falls onto a black hole, collides with the particle $1$. Let's calculate if the unbounded energy $E_{c.m.}$ of the collision is possible and where this collision should take place.

For simplisity, let's consider the collision of two particles with same mass $m_0$. In this case, the energy $E_{c.m.}$ is given by:
\begin{equation} \label{eq:energycm}
E_{c.m.}=m_0\sqrt{2}\sqrt{1-g_{ik}u^i_1u^k_2} \,.
\end{equation}
Where $u^i_1$ and $u^i_2$ are the four velocities of the particles $1$ and $2$ respectivelly.  Substituting \eqref{eq:energy}, \eqref{eq:momentum} and \eqref{eq:radial} into \eqref{eq:energycm}, one obtains:
\begin{equation} \label{eq:fourterms}
\frac{E_{c.m.}^2}{2m_0}=1+\frac{E_1\left (E_2+e^{\frac{2t}{r_0}}P_{2R}\right)}{e^{\frac{2t}{r_0}}\left(1-\frac{2\mu r_0}{R}-\frac{2R}{r_0}\right)}-\frac{e^{\frac{2t}{r_0}}\left(E_1+e^{\frac{2t}{r_0}}P_{1R}\right)P_{2R}}{1-\frac{2\mu r_0}{R}-\frac{2R}{r_0}}-\frac{L_1L_2}{e^{\frac{2t}{r_0}}R^2} \,.
\end{equation}
Note that only the second and third terms in the right-hand side can give us the unbound energy $E_{c.m.}$. It is possible if:
\begin{equation}
1-\frac{2\mu r_0}{R_{kh}}-\frac{2R_{kh}}{r_0}  =0 \,.
\end{equation}
Where $R=R_{kh}$ is the location of the conformal Killing horizon. Also one should note that for outgoing particle $1$ - $P_{1R}>0$, for ingoing particle $2$ - $P_{2R}<0$. So we conclude that both considered terms are positive if we consider the region \eqref{eq:region} of timelike conformal Killing vector $\frac{d}{dt}$ and we should prove that  one of them grows unboundly when $R\rightarrow R_{kh}$. To proceed, we note that:
\begin{equation}
\begin{split}
\lim\limits_{R\rightarrow R_{kh}} e^{\frac{2t}{r_0}} P_{1R}=+E \,, \\
\lim\limits_{R\rightarrow R_{kh}} e^{\frac{2t}{r_0}}P_{2R}=- E \,.
\end{split}
\end{equation}
One should note that off-diagonal term in the metric \eqref{eq:metric2} might indicate that there are particles with negative energy. However, It was shown~\cite{bib:vernegative} that there are not particles with negative energy outside the apparent horizon and $E\geq 0$. Using this fact and by taking limit $R\rightarrow R_{kh}$ one can see that the second term in the right-hand side of the \eqref{eq:fourterms} gives us uncertainty $0/0$ and we won't consider it because if it is finite then we can neglect it. If it is infinite, then we obtain the unbound $E_{c.m.}$. However, we focus our attention on the third term in the right-hand side of \eqref{eq:fourterms}:
\begin{equation} \label{eq:infinity}
\lim\limits_{R\rightarrow R_{kh}} -\frac{e^{\frac{2t}{r_0}}\left(E_1+e^{\frac{2t}{r_0}}P_{1R}\right)P_{2R}}{1-\frac{2\mu r_0}{R}-\frac{2R}{r_0}}=\frac{e^{2t}{r_0}2E_1E_2}{1-\frac{2\mu r_0}{R}-\frac{2R}{r_0}}\rightarrow +\infty \,.
\end{equation}
And we can see that this term \eqref{eq:infinity} gives us the unbound energy $E_{c.m.}$ if the collision takes place in the vicinity of the conformal Killing horizon. 

\section{Conclusion}

In this paper, we have considered the front collision of two particles in the Vaidya spacetime. The metric \eqref{eq:metric} is time-depended and to consider the center-of-mass energy, one needs to introduce new coordinates which allow us to write the Vaidya spacetime in conformally static form. It allows us to consider the following model: in the case of the linear mass function, the gravitational collapse might end up with the naked singularity. We consider the non-spacelike geodesic which originates at this naked singularity. Further, we assume that there are particles which move along this geodesic away from the central singularity. Then, the other particle falls onto a collapsing cloud and the frontal collisions of the two particles is considered at the time of the apparent horizon formation. It means that the apparent horizon forms and the particle, moving along a naked singularity geodesic, finds itself outside the trapped region, in the vicinity of the apparent horizon. We showed that the unbound center-of-mass energy is possible if the collision takes place in the vicinity of the conformal Killing horizon. Note that if we pick up the mass function as $M(v)=\mu v^n\,, n>1$ then, of course, one has the naked singularity formation~\cite{bib:vernon} but the singularity is gravitationally weak according to the Tipler definition~\cite{bib:tip, bib:ir} and the spacetime doesn't admit the conformal Killing vector anymore. 

The unbound center-of-mass energy of two colliding particles near the conformal Killing horizon is expected result. One should use this horizon to define most  physically relevant quantities. In static spacetimes, for example, one uses the Killing horizon to define the surface gravity which is associated with the Hawking temperature. The Killing horizon coincides with the event horizon in the case of Schwarzschild and non-extremal Reissner-Nordstrom black holes. However, in the dynamical case it is not easy task to define the surface gravity~\cite{bib:neldyn}. One can define the surface gravity on the apparent horizon but, according to Nielsen~\cite{bib:nel} the apparent horizon in Vaidya spacetime is hidden inside the event horizon, although, to define the location of the last one is also hard task in dynamical spacetimes. 

The results, obtained in this paper, can be easily extended to generalized Vaidya spacetime. The naked singularity formation and the mass function conditions in this case for this metric has been proven in ~\cite{bib:mah, bib:ver2, bib:ver3}. The unbound center-of-mass is again expected if the front collision takes place in the vicinity of the conformal Killing horizon. However, the conformal Killing vector exists only for the following choice of the mass function:
\begin{equation}
M(v,r)=\mu v+\nu v^{2\alpha} r^{1-2\alpha} \,, \mu>0 \nu \neq 0 \,, \alpha \neq \frac{1}{2}\,.
\end{equation}
Where $\alpha \in [-1\,, 1]$~\cite{bib:vunk}. Note, that for this choice of the mass function, the conformal Killing vector is the homothetic one. The generalized Vaidya spacetime admits regular black hole solution~\cite{bib:regular}, however, the question about the front collision in this case is still open.

\textbf{acknowledgments}: The author says thanks to grant NUM. 22-22-00112 RSF for financial support. The work was performed within the SAO RAS state assignment in the part "Conducting Fundamental Science Research".


\begin{thebibliography}{150}
\bibitem{bib:bsw} M. Banados, J. Silk, S.M. West, Phys. Rev. Lett. 103, 111102 (2009) [arXiv:0909.0169].
\bibitem{bib:zaslavtan} I. V. Tanatarov and O. B. Zaslavskii, Phys. Rev. D 88, 064036 (2013) [arXiv:1307.0034]
\bibitem{bib:zaslavanti} O. B. Zaslavskii, Phys. Lett. B 712 (2012) 161 [arXiv:1202.0565]
\bibitem{bib:japan}    T. Harada, H. Nemoto, U. Miyamoto, Phys.Rev.D86:024027,2012 [arXiv:1205.7088]
\bibitem{bib:zaslavextract} O. B. Zaslavskii, Phys. Rev. D 86, 124039 (2012) [arXiv:1207.5209]
\bibitem{bib:pavlov} A. A. Grib and Yu. V. Pavlov, Gravitation and Cosmology 21 (2015), 13-18 [arXiv:1410.5736]
\bibitem{bib:grib} Grib A.A., Pavlov Yu.V., Vertogradov V.D. Geodesics with negative energy in the ergosphere of rotating black holes. Modern Physics Letters A. Vol. 29,Iss. 20. 2014- P. 14501-14510. [arXiv:1304.7360]
\bibitem{bib:vertog} Vertogradov V.D. Geodesics with negative energy in the ergosphere of rotating black holes. // Gravitation and Cosmology. Vol. 21, Iss. 2. 2015.- PP. 171-174. [arXiv:2210.04674]
\bibitem{bib:vernature} V.D. Vertogradov, The nature of the naked singularity in generalized Vaidya spacetime and white hole geodesics. Physics of Complex Systems, 2021, vol. 2, no. 1
\bibitem{bib:joshi1} Pankaj S. Joshi Gravitational collapse and spacetime singularities. Cambridge University Press. 2007.p.273.
\bibitem{bib:joshi2} Pankaj S. Joshi, D. Malafarina, "Recent development in gravitational collapse and spacetime singularities". Int. J. Mod. Phys. D, 20 2641 (2011)[arXiv:1201.3660]
\bibitem{bib:dey} Dipanjan Dey, Karim Mosani, Pankaj Joshi, Vitalii Vertogradov. Causal structure of singularity in non-spherical gravitational collapse. Eur. Phys. J. C 82, 431 (2022) [arXiv:2103.07190]
\bibitem{bib:dvedy} I H Dwivedi and P S Joshi On the nature of
naked singularities in Vaidya spacetimes Class. Quantum Grav. 6 1599 (1989).
\bibitem{bib:mah}M. D. Mkenyeleye, R. Goswami and S. D. Maharaj,
Phys. Rev. D 90, 064034, (2014).
\bibitem{bib:ver1}Vertogradov V. The eternal naked singularity formation
in the case of gravitational collapse of generalized Vaidya spacetime
Volume No.33, Issue No. 17.
\bibitem{bib:ver2} V.D. Vertogradov Naked singularity formation in
generalized Vaidya space-time. Grav. Cosmol. 22, 220-223 (2016).
\bibitem{bib:ver3} Vertogradov V. Gravitational collapse of Vaidya
spacetime. International Journal of Modern Physics: Conference Series
Vol. 41. 2016. Art. 1660124.
\bibitem{bib:ver4} Vitalii Vertogradov. The structure of the generalized Vaidya spacetime containing the eternal naked singularity. Accepted to international journal of modern physics. [arXiv:2209.10953.]
\bibitem{bib:vay} P. C. Vaidya. Nonstatic solutions of Einstein's field equations for spheres of fluids radiating energy. Phys. Rev., 83:10, 1951.
\bibitem{bib:pap} A. Papapetrou, In: A Random Walk in Relativity and Cosmology. Wiley Eastern, New Delhi (1985).
\bibitem{bib:pois} Poisson E. A Relativist's Toolkit: The Mathematics of Black-Hole Mechanics. Cambridge University Press (2007).
\bibitem{bib:maharaj}   Samson Ojako et al 2020 Class. Quantum Grav. 37 055005 https://arxiv.org/abs/1904.08120v1
\bibitem{bib:nelson} A. B. Nielsen. Revisiting Vaidya horizons. Galaxies, 2:62, 2014.
\bibitem{bib:german} Jay Solanki, Volker Perlick Phys. Rev. D, 105:064056, (2022), arXiv:2201.03274 (gr-qc)
\bibitem{bib:newbook} Matthias Blau Lecture Notes on General Relativity
\bibitem{bib:vernegative}  V. Vertogradov, Universe 6(9), 155 (2020), arXiv:2209.10976 (gr-qc)
\bibitem{bib:vernon} V.D. Vertogradov, Non-linearity of Vaidya spacetime and forces in the central naked singularity. Physics of Complex Systems, 2022, vol. 3, no. 2 [arXiv:2203.05270]
\bibitem{bib:tip} F. J. Tipler, Phys. Lett. A 64, 8 (1977).
\bibitem{bib:ir} B. C. Nolan, Phys. Rev. D 60, 024014 (1999).
\bibitem{bib:neldyn} Nielsen, A.B.; Yoon, J.H. Dynamical surface gravity. Class. Quant. Gravity 2008, 25, 085010, doi:10.1088/0264-9381/25/8/085010.
\bibitem{bib:nel} Nielsen, A.B. The Spatial relation between the event horizon and trapping horizon. Class. Quant. Gravity 2010, 27, 245016, doi:10.1088/0264-9381/27/24/245016.
\bibitem{bib:vunk} A. Wang, Yu.Wu Generalized Vaidya solutions. Gen Relativ. Gravit., 31, 107 (1999)
\bibitem{bib:regular} Sean A. Hayward Formation and Evaporation of
Nonsingular Black Holes Phys. Rev. Lett. 96, 031103
\end{thebibliography}
\end{document}